\documentclass{iopart}
\usepackage{graphicx}
\usepackage{siunitx}
\usepackage{amsmath}
\usepackage{makecell}
\usepackage{epsfig}
\usepackage{array}
\usepackage{braket}
\usepackage[mathscr]{euscript}
\usepackage{multirow}

\begin{document}
\title{Quantified absorption of laser light by silver atoms in a hollow-cathode lamp}

\author{Matthew P. Wilde\textsuperscript\textdagger, Richard G. Wolfendale\textsuperscript\textdagger, \mbox{Mark Bengyel}, \mbox{Ifan G. Hughes}, and Philip D. Gregory}

\address{\textit{Department of Physics, Durham University, South Road, Durham, DH1 3LE, United Kingdom.}}

\footnote[0]{\textsuperscript\textdagger These authors contributed equally}
\ead{p.d.gregory@durham.ac.uk}

\begin{abstract}
We present a quantitative study of laser absorption on the $5{^{2}}\mathrm{S}_{1/2} \rightarrow 5{^{2}}\mathrm{P}_{3/2}$ transition of silver in a hollow-cathode lamp. Spectra obtained using a weak probe are measured and compared to a simple theoretical model. We fit our results to extract an effective temperature and number density associated with the silver in the lamp and confirm that the spectra are dominated by Doppler broadening. We consider the broadening effect of collisions between silver and the neon buffer gas in the lamp, and establish that these effects are small but modify the optimum effective parameters for the model. Our work results in a simple, quantitative, and predictive model for laser absorption in the lamp.  
\end{abstract}

\vspace{2pc}

\maketitle

\section{Introduction}
\label{sec intro}

Thermal atomic vapours offer a simple and robust platform for real-world quantum technologies with applications ranging from quantum sensing~\cite{degen2017quantum, fabricant2023build,  downes2023practical, Brown2025} to the realisation of novel optical filters~\cite{Uhland2023}. Underpinning these applications is a fundamental and quantitative understanding of atom-light interactions that has been developed by the atomic physics community over many decades, with alkali metals generally being the atomic species of choice for experimental simplicity~\cite{pizzey2022laser}. Predictive models for the light absorption of alkali-metal vapours around the $\mathrm{D}$ lines are available in the open-source code {\it ElecSus}~\cite{Zentile2015,Keaveney2018} and have been shown to precisely agree with experimental measurements in potassium~\cite{Hanley2015, alqarni2025role} and rubidium~\cite{Siddons2008}. However, the absorption associated with more exotic species beyond the alkali metals is less well studied. 

Precise spectroscopy of silver (Ag) has important application in astrophysics where determining the relative abundance of isotopes in stars can help constrain the origin of the weak $r$-process~\cite{Jonsson2026, Caliskan2026}. There has also been much recent interest in the laser cooling of Ag~\cite{Uhlenberg2000,Vayninger2025} due to the possibility of creating polar molecules that have large electric dipole moments~\cite{Smialkowski2021}, and which may be useful for precision tests of fundamental physics~\cite{Fleig2021, Klos2022}. The ground electronic structure of Ag is [Kr]~$4\mathrm{d}^{10}~5\mathrm{s}^1$ which, along with the other coinage metals, bears a similarity to that of the alkalis; although it is in group~11 of the periodic table it possesses only a single true valence electron in the highest principal level with the remaining 10 electrons occupying a lower level. The $\mathrm{5s}$ electron in Ag may be resonantly excited with strong one-photon transitions equivalent to the alkali D~lines~\cite{pizzey2022laser}, though with the transition energy shifted to the ultraviolet due to the increased positive charge in the nucleus combined with only weak shielding associated with the filled $4\mathrm{d}$ subshell. Despite this similarity, the study of thermal samples of Ag presents a significant step up in experimental complexity. A major reason is that the vapour pressure is several orders of magnitude smaller than that of the alkalis~\cite{Alcock1984}. Consequently, a vapour cell of Ag would need to be heated to a temperature of $\sim650^{\circ}$C in order to achieve similar vapour pressure to that of an equivalent Rb cell at room temperature. 

One widely-used approach to circumvent this vapour-pressure limitation is the hollow-cathode lamp~\cite{thorne1999spectrophysics}, which provides a convenient source of Ag atoms without the need for extreme heating. The lamp gets its name from a cathode in the shape of a hollow cylinder with an inner surface made from the element under investigation. This cathode, along with a ring-shaped anode, is placed in a glass cell that is filled with an inert buffer gas. A voltage applied across the cathode and anode ionises the inert gas and accelerates the produced ions in the direction of the cathode. Collisions between the ions and the cathode surface eject atoms of the target element. These displaced atoms from the cathode may occupy electronically-excited states initially and their fluorescence is detected as the atoms relax. Inside of the cathode a sufficiently dense vapour forms for laser spectroscopy of the target element. The rate at which atoms are ejected from the cathode surface, and hence the atomic density within the lamp, is strongly dependent on the discharge current flowing through the plasma. 

Hollow-cathode lamps are commercially available and have found a range of applications. The emitted fluorescence from hollow-cathode lamps is routinely used in atomic absorption spectrometers~\cite{Beaty1993} and as a frequency calibrator for many high resolution astronomical spectrographs~\cite{sarmiento2018comparing}. Hollow-cathode discharge can be used as the gain medium for a laser~\cite{Schuebel1970}, and laser absorption in the lamp near resonance may be detected by changes in the current flowing through the lamp in a technique known as optogalvanic spectroscopy~\cite{Barbieri1990}. Direct measurements of laser transmission through hollow cathode lamps have been reported for a number of elements in the context of laser frequency stabilisation~e.g.~\cite{Cavasso-Filho2001,Smeets2003,Kim2003, Aoki2012,Chen2013, liu2018ultraviolet, Angonga2018, blums2020laser}, optical filtering~\cite{Pan2016,Shen2020, luo2021thermal}, or for precision measurement~\cite{Liu2018}. Laser spectroscopy of Ag in hollow-cathode lamps in particular was the subject of a series of investigations to spectroscopically study the metastable $4\mathrm{d}^9~5\mathrm{s}^2$ states and their associated transitions for application in atomic clocks~\cite{Guerandel2000, Badr2001,Badr2004,Badr2006}.

In this article, we measure weak-probe absorption on the $5{^{2}}\mathrm{S}_{1/2} \rightarrow 5{^{2}}\mathrm{P}_{3/2}$ transition of Ag in a hollow cathode lamp. We compare our results to a theoretical model and find that the observed absorption profile can be quantitatively understood by a simple combination of the natural linewidth of the transition and Doppler broadening. By fitting the measured absorption spectra, we extract coefficients relating the Ag number density and effective temperature in the lamp to the discharge current.

The remainder of the article is structured as follows: In section 2, we outline our model for absorption in the silver hollow cathode lamp. In section 3, we give details of our experimental apparatus. In section 4, we fit and compare our model to the measured absorption and extract effective parameters for the temperature and number density in the lamp. 

\section{Modelling the absorption in a silver hollow cathode lamp}

Our model is similar to the treatment of atomic absorption used in {\it ElecSus} for thermal vapours~\cite{Zentile2015}, but crucially we break any link between the temperature of the gas and the density of the atoms. We assume that the silver atoms form a column of uniform density with length equal to that of the cathode, i.e. $l=20$\,mm. The fraction of light transmitted through the medium is then given by the Beer-Lambert law
\begin{equation}
\mathcal{T} = \exp{\left(-\alpha l\right)}.
\label{eq:BeerLambert}
\end{equation}
Here, $\alpha$ is the absorption coefficient that is related to the imaginary component of the electronic susceptibility $\chi_i$ of the medium by $\alpha = k\chi_i$ where $k=2\pi/\lambda$ is the wavenumber of the probe beam with wavelength $\lambda$.  We note that we plot all of our results and calculations as a function of linear frequency; a factor of $2\pi$ converts between angular (rad\,s$^{-1}$) and linear (Hz) frequency units~\cite{Mohr2015}.

The absorption coefficient generally depends on the frequency of the light ($\nu$) relative to the resonant frequency of the atomic transition ($\nu_0$), and the temperature ($T$) and number density ($n$) of the atoms, while its exact form depends on the relative contributions of homogeneous and inhomogeneous broadening mechanisms.  Homogeneous mechanisms include spontaneous emission from the excited state resulting in the natural linewidth $\Gamma=23.4$\,MHz~\cite{Bengtsson1990} and pressure-broadening associated with velocity-independent collisions with the inert buffer gas~\cite{Foley1946}. These result in a Lorentzian lineshape 
\begin{equation}
\mathcal{L}=\frac{1}{\pi}\frac{\Gamma_\mathrm{tot}/2}{(\nu-\nu_0)^2+(\Gamma_\mathrm{tot}/2)^2}.
\end{equation}
Here, $\Gamma_\mathrm{tot}=\Gamma+\Gamma_\mathrm{self}+\Gamma_\mathrm{buf}$ where we include self-broadening governed by~$\Gamma_\mathrm{self}=\sqrt{2}\Gamma n(\lambda/2\pi)^3$~\cite{lewis1980collisional, weller2011absolute}, and $\Gamma_\mathrm{buf}$ characterises the broadening due to the collisions with the buffer gas which is generally temperature dependent~\cite{alqarni2025role}. The dominant inhomogeneous mechanism in the hollow-cathode lamp is assumed to be Doppler broadening. This corresponds to a Gaussian lineshape of the form
\begin{equation}
\mathcal{G}=\sqrt{\frac{m}{2\pi k_\mathrm{B} T}}\exp\left[-\frac{m c^2}{2k_\mathrm{B}T}\left( \frac{\nu-\nu_0}{\nu_0}\right)^2 \right],
\end{equation}
where $m$ is the mass of the atom, $c$ is the speed of light, and $k_\mathrm{B}$ is the Boltzmann constant. For a transition broadened by a combination of homogeneous and inhomogeneous mechanisms, the overall lineshape follows a Voigt profile, given by the convolution~\cite{thorne1999spectrophysics}
\begin{equation}
\mathcal{V} = \mathcal{L} \ast \mathcal{G},
\end{equation}
which is used in our model. This generalised lineshape is related to the absorption coefficient for a given individual transition by
\begin{equation}
\alpha=\frac{\pi c_{FF'}^2 d^2 n}{2(2I+1)\lambda\hbar\epsilon_0}\mathcal{V}.
\label{eq:AbsorptionCoeff}
\end{equation}
Here, $c_{FF'}$ and $d$ are the transition strength coefficient and reduced dipole matrix element respectively, $\hbar$ is the reduced Planck constant, $\epsilon_0$ is the permittivity of free space, and $I$ is the nuclear spin such that $(2J+1)(2I+1)=2(2I+1)$ denotes the number of ground states of the atom where $J=L+S$ is the combination of orbital angular momentum and intrinsic spin of the electron.  

\begin{figure}[t!]
\includegraphics[width=\columnwidth]{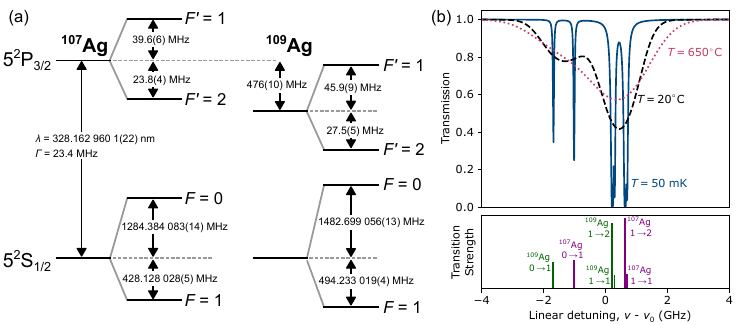}
\caption{(a)~Schematic of the energy levels relevant to the $5{^{2}}\mathrm{S}_{1/2} \rightarrow 5{^{2}}\mathrm{P}_{3/2}$ transition in silver. The transition wavelength is from~\cite{Pickering2001}, the energy difference between ground states was measured in~\cite{Dahmen1967} and the excited states in~\cite{Carlsson1990}, and the isotope shift is taken from~\cite{Uhlenberg2000}. (b)~The top panel shows absorption spectra predicted by our model for silver vapour in the absence of buffer gas broadening at temperatures of -273.10$^{\circ}$C (50\,mK), 20$^\circ$C, and 650$^\circ$C. For illustrative purposes, the number density at $2\times10^{16}$\,m$^{-3}$ as the temperature is varied. The bottom panel shows the relative transition strengths associated with each feature, labelled by the isotope and quantum numbers $F\rightarrow F'$ of the transition.}
\label{fig:Theory}
\end{figure}

Combining Eq.\,\ref{eq:BeerLambert} and Eq.\,\ref{eq:AbsorptionCoeff}, and summing over the available atomic transitions and isotopes, yields a quantitative model for absorption in a silver vapour. We assume that the population is distributed equally among all $2(2I+1)$ ground states.  Silver has two naturally occurring isotopes, each with nuclear spin $I=1/2$ and natural abundance $^{107}\mathrm{Ag}:~51.84\%$ and $^{109}\mathrm{Ag}:~48.16\%$~\cite{Meija2016}. For the $5{^{2}}\mathrm{S}_{1/2} \rightarrow 5{^{2}}\mathrm{P}_{3/2}$ transition under investigation $\lambda=328.16$\,nm, and the relevant energy level diagram is shown in Fig.\,\ref{fig:Theory}(a). We label each state by the total angular momentum quantum number $F$, obtained from the vector coupling of the nuclear and electronic angular momenta $F=I+J$. The reduced dipole matrix element for this transition is 
\begin{equation}
d=\bra{L}\hat{d}\ket{L'}=3\sqrt{\frac{\epsilon_0\hbar\Gamma\lambda^3}{8\pi^2}}=2.35\times10^{-29}\,\mathrm{C\,m}, 
\end{equation}
which is independent of the hyperfine quantum numbers. The relative strengths of individual hyperfine transitions $F\rightarrow F'$ are determined by the transition strength coefficients~\cite{sobelman2012atomic}
\begin{equation}
c_{FF'} = \sqrt{(2F+1)(2F'+1)(2J+1)(2J'+1)(2L+1)}
\begin{Bmatrix}
J & J' & 1 \\
F' & F & I
\end{Bmatrix}
\begin{Bmatrix}
L & L' & 1 \\
J' & J & S
\end{Bmatrix},
\label{eq:cF}
\end{equation}
where the last two terms are Wigner 6-$j$ symbols. For our work the relevant values are $c_{01}=\sqrt{2/3}, c_{11}=\sqrt{1/3}, c_{12}=\sqrt{5/3}$, which we calculate from Eq.\,\ref{eq:cF} with $J=1/2, J'=3/2, L=0, L'=1, I=1/2, S=1/2$. We note that the present work is performed in the absence of a magnetic field. We however include the transition strength coefficients needed with a magnetic field applied in the basis $(F,m_F,F',m_{F'})$ in the supplementary material for completeness.

Example spectra predicted by the model are shown in Fig.\,\ref{fig:Theory}(b) for a range of temperatures and fixed number density. At 50\,mK the individual hyperfine transitions are fully resolved. Increasing the temperature leads to substantial Doppler broadening, causing neighbouring transitions to merge. At room temperature the transitions originating from the $F=0$ and $F=1$ ground states form two partially-overlapping absorption features separated by a shallow local maximum. At still higher temperatures these features are no longer distinguishable and the spectrum approaches a single broad Gaussian-like profile.

\section{Experimental details}

\begin{figure}[t!]
\includegraphics[width=1.02\columnwidth]{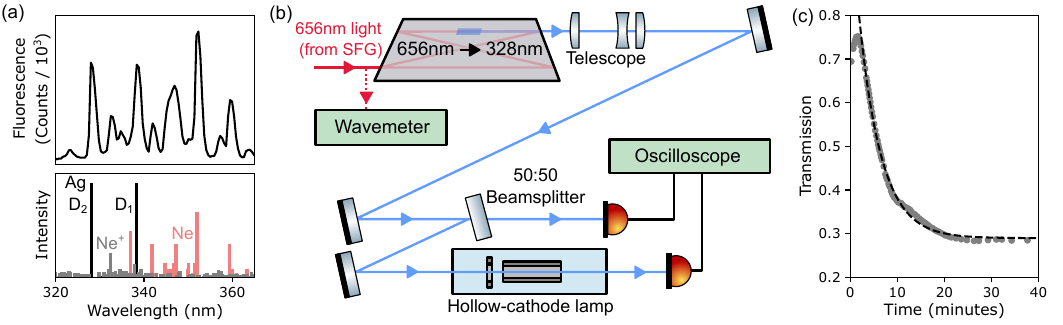}
\caption{(a)~Fluorescence spectrum emitted from the lamp. The results in the top panel were captured by coupling light from the lamp into an optical fibre which was then detected using a commercial spectrometer (Ocean Optics Flame). The bottom panel indicates the expected lines for Ag (black), Ne (red) and ionised Ne$^{+}$ (grey) using values taken from the NIST atomic spectra database~\cite{NIST}. The $5{^{2}}\mathrm{S}_{1/2} \rightarrow 5{^{2}}\mathrm{P}_{3/2}$ transition of silver that is the subject of this study is labelled as Ag~$\mathrm{D}_2$. (b)~Optical apparatus used for laser spectroscopy as described in the main text. Some mirrors have been omitted for clarity. (c)~The minimum transmission through the lamp as a function of time from the lamp being first switched on to a current of~8\,mA.  }
\label{fig:Apparatus}
\end{figure}

We use a commercial see-through hollow-cathode lamp (Spectrolamps HCO52ST) which contains a neon buffer gas of unknown partial pressure. The windows of the lamp are made from quartz glass, and the cylindrical cathode inside is $20$\,mm long and has an inside diameter of approximately $3$\,mm. We run the lamp at the manufacturer-specified operating current ranging from 4\,mA to 8\,mA. Using a spectrometer we have detected fluorescence from neutral neon, ionised neon, and silver across this range of currents; an example fluorescence spectrum in the region around the silver line of interest is shown in Fig.\,\ref{fig:Apparatus}(a).

For spectroscopy we generate laser light at a wavelength of 328\,nm using a combination of sum-frequency generation and second-harmonic generation, starting with near-infrared light with wavelengths of 1595\,nm and 1114\,nm. The 1595\,nm light is generated by a fixed-external-cavity laser which seeds an 8\,W fibre amplifier (Precilaser FECL-SF-1595-S and FA-SF-1595-8-CW), and the 1114\,nm light by a distributed-feedback laser which in turn seeds a 15\,W fibre amplifier (Precilaser FL-SF-1114-S and FA-SF-1114-15-CW). The light from both amplifiers are combined on a dichroic mirror and focused into a periodically-poled nonlinear crystal (MgO:PPLN, Covesion MSFG647-0.5-40) that is heated and stabilised at a temperature of 174$^{\circ}$C to generate 656\,nm light via single-pass sum-frequency generation. The 656\,nm light is separated from the near-infrared light using dichroic mirrors and coupled to a photonic crystal fibre (NKT photonics aeroGUIDE POWER). The output of the fibre is coupled to a bow-tie enhancement cavity (Agile Optics) for second harmonic generation using cesium lithium borate (CLBO). The CLBO is kept heated to a temperature of $140^\circ$C and the cavity continuously purged with dry air to prevent degradation. This system is capable of generating in excess of 100\,mW of ultraviolet light at 328\,nm for magneto-optical trapping of silver in a separate ultra-high vacuum apparatus~\cite{Bengyel2026}.

Our optical setup after the bow-tie enhancement cavity is shown in Fig.\,\ref{fig:Apparatus}(b). The 328\,nm light output from the cavity is linearly-polarised vertically, and initially highly elliptical and astigmatic. We use cylindrical lenses in a telescope to shape the beam; at the position of the hollow-cathode lamp we achieve $1/\mathrm{e}^2$ beam radii of $\omega_\mathrm{h}=0.87(3)$\,mm and $\omega_\mathrm{v}=0.92(3)$\,mm in the horizontal and vertical directions respectively, with corresponding half-angle divergence $\theta_\mathrm{h}/2=0.11(2)$\,mrad and $\theta_\mathrm{v}/2=0.15(3)$\,mrad. We use a series of beam samplers and glass wedges to reduce the average power of the beam at the hollow-cathode lamp to 15(1)\,\textmu W. 

To measure laser absorption in the lamp while eliminating effects from noise in the laser intensity, we sample the beam both before and after it passes through the hollow-cathode lamp using a 50:50 beamsplitter and two photodiodes; we simultaneously record the signals from both photodiodes on an oscilloscope and convert to transmission by dividing them. We use coloured glass filters to minimise background room light from reaching the photodiodes (Thorlabs FGUV). In addition, both photodiodes are placed far from the lamp, such that there is negligible signal from the lamp's fluorescence. The saturation intensity of the $5{^{2}}\mathrm{S}_{1/2} \rightarrow 5{^{2}}\mathrm{P}_{3/2}$ transition is $I_\mathrm{sat}=87$\,mW\,cm$^{-2}$, such that the peak intensity of the beam corresponds to~\mbox{$\approx 0.014 I_\mathrm{sat} $}; this is sufficiently low to be considered in the weak-probe regime where the absorption does not depend on the laser power~\cite{Sherlock2009}. 

We find that the absorption in the lamp requires time to stabilise after turning on. A measurement of peak absorption in the lamp as a function of time after first switching the lamp on to 8\,mA is shown in Fig.~\ref{fig:Apparatus}(c). The transmission briefly increases slightly in the first 5\,minutes after which the transmission decreases until it reaches a steady state. We fit the absorption after these first 5 minutes with an exponential function and find the $1/e$ time constant for this behaviour to be~4.72(1)\,minutes. We therefore ensure that the lamp is switched on at the desired current for at least 30\,minutes prior to performing any measurements. 

The detuning of the 328\,nm light was varied by changing the frequency of the 1595\,nm seed laser via the laser current. All frequency measurements were performed on the intermediate 656\,nm light using a wavemeter (Bristol 621A), and subsequently doubled to obtain the probe frequency at 328\,nm.  The manufacturer-specified absolute accuracy of the wavemeter is $\pm90$\,MHz at the $3\sigma$ confidence level. By using the wavemeter to monitor a frequency-stabilised reference laser, we estimate a short-term relative frequency resolution of $\sim10$\,MHz relevant to our measurements. To perform spectroscopy, we used computer control to automatically set and iterate the 1595\,nm laser frequency, measure and average the frequency of the 656\,nm light over several seconds, and simultaneously record oscilloscope traces from each of the photodiodes. Each complete spectrum is captured in around 30\,minutes using this process.

\section{Results and discussion}

\begin{figure}[t!]
\includegraphics[width=\columnwidth, trim={0.5cm, 0cm, 0.4cm, 0cm}]{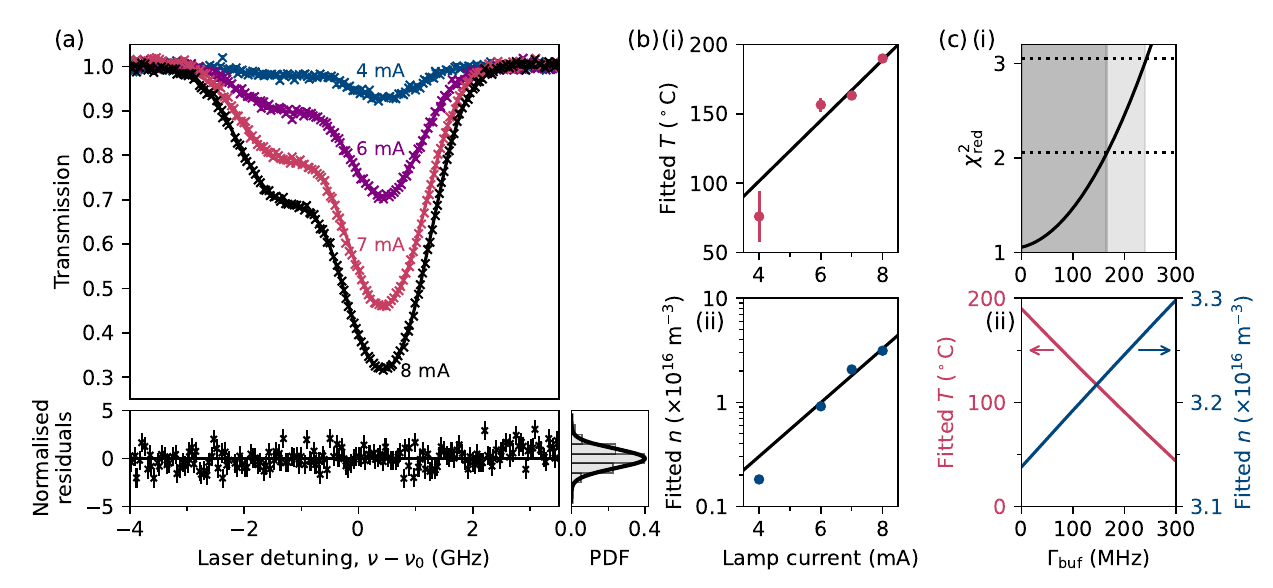}
\caption{(a)~Experimentally measured spectra at lamp currents of 4-8\,mA. Fits to the results are shown by solid lines with $\Gamma_\mathrm{buf}=0$. The normalised residuals associated with the fit to the 8\,mA lamp current are shown in the bottom panel alongside a histogram of their probability density function (PDF). The solid line on the histogram indicates the expected normal distribution. (b)~Variation in the fitted parameters as a function of the lamp current. (i)~The effective temperature of the silver vapour, with a linear fit shown by the solid line. (ii)~The number density with an exponential fit shown by the solid line; note that the $y$-axis is on a logarithmic scale. (c)~The effect of adding buffer-gas broadening to the fit. (i)~The reduced chi-squared $\chi_\mathrm{red}^2$ achieved in fitting the 8\,mA lamp current results as a function of the buffer gas broadening included in the model, which is characterised by the Lorentzian width $\Gamma_\mathrm{buf}$. The dark-grey and light-grey shaded regions indicate the $1\sigma$ and $2\sigma$ confidence intervals respectively. (ii)~The effect on the optimal values of the temperature $T$ and number density $n$ found in the fitting as function of the buffer gas broadening.  }
\label{fig:result}
\end{figure}

Experimentally measured spectra at lamp currents between 4~and~8\,mA are shown in~Fig.\,\ref{fig:result}(a). We fit each spectrum with our model assuming no buffer gas broadening; the free parameters in the fit are the effective temperature $T$, the number density $n$, and a frequency offset. The lack of structure in the residuals indicates excellent agreement between our theoretical model and the experimental data~\cite{hughes2010measurements}. The Voigt lineshapes we observe are closer in character to a Gaussian than a Lorentzian as expected of a system dominated by Doppler broadening.

The optimum parameters found in the fitting are shown in Fig.\,\ref{fig:result}(b). The fitted effective temperatures are substantially lower than those that would be required to generate comparable absorption in a conventional thermal vapour, and low enough to discern that there are separate transitions associated with the $5^2\mathrm{S}_{1/2}$~$F=0$~and~$F=1$ ground states. This highlights the ability of the hollow-cathode discharge to produce high atomic densities without heating the sample to extreme temperatures. When running the lamp at maximum current, we estimate that the outside surface of the lamp closest to the hollow cathode heats up to a temperature of $\sim 47^\circ$C based on measurements with a thermocouple taped to the outside surface of the glass. We perform a linear fit to the temperature data and estimate a gradient of~$22(3)^\circ\mathrm{C}\,\mathrm{mA}^{-1}$ and $y$-intercept equal to~$14(17)^\circ\mathrm{C}$. We fit to the variation in the number density using an exponential of the form $n=A\mathrm{e}^{kI_\mathrm{lamp}}$ where $I_\mathrm{lamp}$ is the lamp current, and find optimum parameters of $A=2.78(3)\times10^{14}\,\mathrm{m}^{-3}$, $k=0.597(2)\,\mathrm{mA}^{-1}$. These fitted trends allow us to make quantitative predictions of behaviours between the lamp currents that we have investigated. 

So far, we have neglected the role of the lamp's neon buffer gas in setting the width of the spectral lines. We therefore investigate the effect of adding buffer-gas broadening to our model on the fitting of the 8\,mA result in Fig.\,\ref{fig:result}(c) by setting non-zero values of $\Gamma_\mathrm{buf}$. This essentially increases the weight of the Lorentzian component present in the absorption profile. We see that any amount of broadening associated with the buffer gas reduces the quality of the fit, which is indicated by an increase in the reduced chi-squared. However, the confidence interval is broad, such that our measurements would be compatible with values of $\Gamma_\mathrm{buf}<240$\,MHz at the $2\sigma$ confidence level.

The  magnitude of the buffer-gas broadening depends upon the interatomic potentials for Ag+Ne for silver prepared in the $5{^{2}}\mathrm{S}_{1/2}$ and $5{^{2}}\mathrm{P}_{3/2}$ states. These have been  calculated previously by Loreau~{\it et~al.}~\cite{Loreau2013}, though no broadening coefficients  have been estimated. From the potentials we coarsely estimate a broadening coefficient of the order of $\sim10$\,MHz\,Torr$^{-1}$ based on the  difference in the ground state and excited state Ag-Ne potentials. Here, we fitted only the long-range parts of the calculated potentials (interatomic separation $>12\,\mu$m) with a much simpler potential of the form $C_6/r^6$ , where $r$ is the interatomic separation and $C_6$ is a constant. We then calculate the broadening coefficient using the expression given in Eq.\,(27c) of~\cite{Mizushima1951}, which is derived from an adiabatic impact theory approximation. We note that our analysis is likely more valid for collisions associated with the $\mathrm{A}\,^3\mathrm{\Pi}_{3/2}$ excited potential that is purely attractive, than for the $\mathrm{B}\,^2\mathrm{\Sigma}_{1/2}$ potential that exhibits a shallow well that is not captured by our simplified potential. For comparison, buffer-gas broadening coefficients have been experimentally measured for alkali atoms in the presence of neon; the exact value depends on the atomic species, choice of transition, and the temperature, but for the $\mathrm{D}_2$ line of K, Rb, and Cs the value is around 2\,MHz\,Torr$^{-1}$~\cite{Pitz2014}. The typical operating pressure of hollow cathode lamps is in the range of 1-10\,Torr~\cite{Barbieri1990}. This would suggest buffer-gas broadening in the lamp is likely comparable to the natural linewidth of the transition, therefore modifying the Lorentzian contribution to the Voigt, but this effect is relatively small in the presence of such a large Doppler profile. 

We examine the effect of introducing the buffer gas broadening on the fit parameters in~Fig.\,\ref{fig:result}(c)(ii). We find that the optimum values of $T$ and $n$ both vary linearly with $\Gamma_\mathrm{buf}$ though with opposite sign, and $T$ is much more sensitive to the inclusion of the buffer-gas broadening than $n$. We find the fitted value of $T$ varies at a rate of -0.4799(5)$^{\circ}$C\,MHz$^{-1}$ and $n$ varies as~$5.303(3)\times10^{12}$\,m$^{-3}$\,MHz$^{-1}$. It is therefore straightforward to include buffer-gas broadening in the absorption model, and its effect is to modify the effective temperature and number density.

In conclusion, we have presented laser absorption spectroscopy of the $5{^{2}}\mathrm{S}_{1/2} \rightarrow 5{^{2}}\mathrm{P}_{3/2}$ transition of silver. We have adapted an existing theoretical framework to analyse these results and shown that excellent agreement between theory and experiment can be achieved with a relatively simple model that is governed by an effective temperature and number density in the lamp. The inclusion of buffer-gas broadening results in a modification of these effective parameters. Although the present measurements do not exclude a contribution from Ag-Ne collisions, they indicate that any additional broadening is not required to reproduce the observed spectra within the experimental uncertainty. Our work paves the way for many of the techniques developed for alkali thermal vapours to be extended to more exotic species through the use of hollow-cathode lamps, where a quantitative and predictive model for the absorption will be an invaluable tool.

\section*{Acknowledgements}

PDG is supported by a Royal Society University Research Fellowship (URF/R1/231274) and the UK Engineering and Physical Sciences Research Council (EPSRC) UKRI2226 funded through the Programme Grant Scheme. MB is supported by a UKRI quantum technologies studentship. We thank J\'{e}r\^{o}me Loreau for supplying the numerical values for the potentials in~\cite{Loreau2013} and Danielle Pizzey for initial set up of the hollow cathode lamp, the loan of the spectrometer, and for helpful comments on the manuscript. 

\section*{Data availability}
All data presented in this manuscript are available at DOI:10.15128/r19880vr083.

\section*{Code availability}
The codes used in this manuscript are available in the GitHub repository: \\ github.com/PhilipDGregory/Quantified-absorption-of-laser-light-by-silver-atoms-in-a-hollow-cathode-lamp. 

\section*{References}
\bibliographystyle{unsrt}
\bibliography{ref}

@article{Brown2025,
  title = {Perspective: Practical atom-based quantum sensors},
  author = {Brown, Justin M. and Walker, Thad G.},
  journal = {Phys. Rev. A},
  volume = {112},
  issue = {4},
  pages = {040102},
  numpages = {33},
  year = {2025},
  month = {Oct},
  publisher = {American Physical Society},
  doi = {10.1103/q6d1-594v},
  url = {https://link.aps.org/doi/10.1103/q6d1-594v}
}

@article{Uhland2023,
  title = {How to build an optical filter with an atomic vapor cell},
  author = {Uhland, Denis and Dillmann, Helena and Wang, Yijun and Gerhardt},
  journal = {New J. Phys.},
  volume = {25},
  pages = {125001},
  year = {2023},
  doi = {10.1088/1367-2630/ad0fa8},
}

@article{Zentile2015,
  title = {{ElecSus:} {A} program to calculate the electric susceptibility of an atomic ensemble},
  author = {Zentile, Mark A. and Keaveney, James and Weller, Lee and Whiting, Daniel J. and Adams, Carles S. and Hughes, Ifan G.},
  journal = {Comp. Phys. Commun.},
  volume = {189},
  pages = {162-174},
  year = {2015},
  doi = {10.1016/j.cpc.2014.11.023},
}

@article{Keaveney2018,
  title = {{ElecSus:} {E}xtension to arbitrary geometry magneto-optics},
  author = {Keaveney, James and Adams, Charles S. and Hughes, Ifan G.},
  journal = {Comp. Phys. Commun.},
  volume = {224},
  pages = {311-324},
  year = {2018},
  doi = {10.1016/j.cpc.2017.12.001},
}

@article{Hanley2015,
  title = {Absolute absorption on the potassium {D} lines: theory and experiment},
  author = {Hanley, Ryan K. and Gregory, Philip D. and Hughes, Ifan G. and Cornish, Simon L.},
  journal = {J. Phys. B: At. Mol. Opt. Phys.},
  volume = {48},
  pages = {195004},
  year = {2015},
  doi = {10.1088/0953-4075/48/19/195004},
}

@article{Siddons2008,
  title = {Absolute absorption on the rubidium {D} lines: comparison between theory and experiment},
  author = {Siddons, Paul and Adams, Charles S. and Ge, Chang and Hughes, Ifan G.},
  journal = {J. Phys. B: At. Mol. Opt. Phys.},
  volume = {41},
  pages = {155004},
  year = {2008},
  doi = {10.1088/0953-4075/41/15/155004},
}

@article{Caliskan2026,
  title = {Ag {I} atom and the {3D} non-{LTE} solar silver abundance},
  author = {Caliskan, S. and Amarsi, A. M. and J\"{o}nsson, P. and Grevesse, N. and Sahoo, B. K.},
  journal = {A\&A},
  volume = {711},
  pages = {A155},
  year = {2026},
  doi = {10.1051/0004-6361/202659578},
}

@article{Jonsson2026,
  title = {Accurate transition and hyperfine data in {Ag} {I} from the multiconfiguration {Dirac-Hartree-Fock} and relativistic coupled-cluster methods},
  author = {J\"{o}nsson, Per and Sahoo, B. K. and Caliskan, Sema and Amarsi, A. M.},
  journal = {A\&A},
  volume = {709},
  pages = {A31},
  year = {2026},
  doi = {10.1051/0004-6361/202558824},
}

@article{Uhlenberg2000,
  title = {Magneto-optical trapping of silver atoms},
  author = {Uhlenberg, G. and Dirschel, J. and Walther, H.},
  journal = {Phys. Rev. A},
  volume = {62},
  pages = {063404},
  year = {2000},
  doi = {10.1103/PhysRevA.62.063404},
}

@article{Vayninger2025,
  title = {Magneto-optical trap of silver and potassium atoms},
  author = {Vayninger, Michael and Xiang, Angela and Bhanushali, Nachiket D. and Chen, Xiaoyu and Verma, Mohit and Yang, Shaozhen and Kapur, Rohan T. and DeMille, David and Yan, Zoe Z.},
  journal = {Phys. Rev. A},
  volume = {112},
  pages = {063306},
  year = {2025},
  doi = {10.1103/p974-js81},
}

@article{Smialkowski2021,
  title = {Highly polar molecules consisting of a copper or silver atom interacting with an alkali-metal or alkaline-earth-metal atom},
  author = {\ifmmode \acute{S}\else \'{S}\fi{}mia\l{}kowski, Micha\l{} and Tomza, Micha\l{}},
  journal = {Phys. Rev. A},
  volume = {103},
  pages = {022802},
  year = {2021},
  doi = {10.1103/PhysRevA.103.022802},
}

@article{Fleig2021,
  title = {Theoretical aspects of radium-containing molecules amenable to assembly from laser-cooled atoms for new physics searches},
  author = {Fleig, Timo and De{M}ille, David},
  journal = {New J. Phys.},
  volume = {23},
  pages = {113039},
  year = {2021},
  doi = {10.1088/1367-2630/ac3619},
}

@article{Klos2022,
  title = {Prospects for assembling ultracold radioactive molecules from laser-cooled atoms},
  author = {K\-{l}os, Jacek and Li, Hui and Tiesinga, Eite and Kotochigova, Svetlana},
  journal = {New J. Phys.},
  volume = {24},
  pages = {025005},
  year = {2022},
  doi = {10.1088/1367-2630/ac50ea},
}

@article{Meija2016,
  url = {https://doi.org/10.1515/pac-2015-0503},
  title = {Isotopic compositions of the elements 2013 ({IUPAC} {T}echnical {R}eport)},
  author = {Meija, J. and Coplen, T. B. and Berglund, M. and Brand, W. A. and De Bièvre, P. and Gröning, M. and Holden, N. E. and Irrgeher, J. and Loss, R. D. and Walczyk, T. and Prohaska, T.},
  pages = {293--306},
  volume = {88},
  journal = {Pure Appl. Chem.},
  doi = {doi:10.1515/pac-2015-0503},
  year = {2016},
}

@article{Barbieri1990,
  title = {Optogalvanic spectroscopy},
  author = {Barbieri, Beniamino and Beverini, Nicol\'{o}},
  journal = {Rev. Mod. Phys.},
  volume = {62},
  pages = {603},
  year = {1990},
  doi = {10.1103/RevModPhys.62.603},
}

@book{Beaty1993,
  author = {Beaty, Richard D. and Kerber, Jack D.},
  year = {1993},
  title = {Concepts, instrumentation and techniques in atomic absorption spectrophotometry},
  publisher = {The {Perkin-Elmer} Corporation},
  edition = {2}
}

@article{Schuebel1970,
  author = {Schuebel, Wolfgang K.},
  title = {New cw {Cd}-vapor laser transitions in a hollow-cathode structure},
  journal = {Appl. Phys. Lett.},
  volume = {16},
  pages = {470},
  year = {1970},
  doi = {10.1063/1.1653070},
}

@article{Aoki2012,
  author = {Aoki, Takatoshi and Umezawa, Kotaro and Yamanaka, Yuki and Takemura, Naotomo and Sakemi, Yasuhiro and Torii, Yoshio},
  title = {A 461\,nm laser system and hollow-cathode lamp spectroscopy for magneto-optical trapping of {Sr} atoms},
  journal = {J. Phys. Soc. Jpn.},
  volume = {81},
  pages = {034401},
  year = {2012},
  doi = {10.1143/JPSJ.81.034401},
}

@article{Cavasso-Filho2001,
  author = {Cavasso-{F}ilho, R. L. and Mirage, A. and Scalabrin, A. and Pereira D. and Cruz, F. C.},
  title = {Laser spectroscopy of calcium in hollow-cathode discharges},
  journal = {J. Opt. Soc. Am. B},
  volume = {18},
  pages = {1922},
  year = {2001},
  doi = {10.1364/JOSAB.18.001922},
}

@article{Smeets2003,
  author = {Smeets, B. and Bosch, R. C. M. and {v}an der {S}traten, P. and {te Sligte}, E. and Scholten, R. E. and Beijerinck, H. C. W. and {van} {L}eeuwen, K. A. H.},
  title = {Laser frequency stabilization using an {Fe-Ar} hollow cathode discharge cell},
  journal = {Appl. Phys. B},
  volume = {76},
  pages = {815-819},
  year = {2003},
  doi = {10.1007/s00340-003-1228-1},
}

@article{Chen2013,
  author = {Chen, Tzu-{L}ing and Lin, Chang-{Y}i and Shu, Jow-{T}song and Liu, Yi-{W}ei},
  title = {Tunable frequency-stabilisation of an ultraviolet laser using a hollow cathode lamp of atomic thallium},
  journal = {J. Opt. Soc. Am. B},
  volume = {30},
  pages = {2966},
  year = {2013},
  doi = {10.1364/JOSAB.30.002966},
}

@article{Kim2003,
  author = {Kim, Jae Ihn and Park, Chang Yong and Yeom, Jin Yong and Kim, Eok Bong and Yoon, Tai, Hyun},
  title = {Frequency-stabilized high-power violet laser diode with an ytterbium hollow-cathode lamp},
  journal = {Opt. Lett.},
  volume = {28},
  pages = {245},
  year = {2003},
  doi = {10.1364/OL.28.000245},
}

@article{Angonga2018,
  author = {Ang'ong'a, Jackson and Gadway, Bryce},
  title = {Polarization spectroscopy of atomic erbium in a hollow cathode lamp},
  journal = {J. Phys. B. At. Mol. Opt. Phys.},
  volume = {51},
  pages = {045003},
  year = {2018},
  doi = {10.1088/1361-6455/aaa1d4},
}

@article{Liu2018,
  author = {Liu, Hongli and Yuan, Wenhao and Xu, Zetian and Deng, Ke and Lu, Zehuang},
  title = {Ultraviolet laser spectroscopy of aluminium atoms in hollow-cathode lamp},
  journal = {J. Phys. B. At. Mol. Opt. Phys.},
  volume = {51},
  pages = {225002},
  year = {2018},
  doi = {10.1088/1361-6455/aada9a},
}

@article{Alcock1984,
  author = {Alcock, C. B. and Itkin, V. P. and Jorrigan, M. K.},
  title = {Vapour pressure equations for the metallic elements: {298-2500\,K}},
  journal = {Can. Metall. Q},
  volume = {23},
  pages = {309-313},
  year = {2013},
  doi = {10.1179/cmq.1984.23.3.309},
}

@article{Shen2020,
  author = {Shen, Liang and Ma, Rui and Yin, Longfei and Luo, Bin and Pan, Duo and Yu, Song and Chen, Jongbiao and Guo, Hong},
  title = {A {F}araday anomalous dispersion optical filter based on a rubidium hollow-cathode lamp},
  journal = {Appl. Sci.},
  volume = {10},
  pages = {7075},
  year = {2020},
  doi = {10.3390/app10207075},
}

@article{Pan2016,
  author = {Pan, Duo and Xue, Xiaobo and Shang, Haosen and Luo, Bin and Chen, Jingbiao and Guo, Hong},
  title = {Hollow cathode lamp based {F}araday anomalous dispersion optical filter},
  journal = {Sci. Rep.},
  volume = {6},
  pages = {29882},
  year = {2016},
  doi = {10.1038/srep29882},
}

@article{Guerandel2000,
  author = {Gu\'{e}randel, S. and Badr, T. and Plimmer, M. D. and Juncar, P. and Himbert, M. E.},
  title = {Frequency measurement, isotope shift and hyperfine structure of the $4\mathrm{d}^95\mathrm{s}^2~^2\mathrm{D}_{5/2} \rightarrow 4\mathrm{d}^{10}6\mathrm{p}~^2\mathrm{P}_{3/2}$ transition in atomic silver},
  journal = {Eur. Phys. J. D},
  volume = {10},
  pages = {33-38},
  year = {2000},
  doi = {10.1007/s100530050524},
}

@article{Badr2001,
  author = {Badr, T. and Gu\'{e}randel, S. and Plimmer, M. D. and Juncar, P. and Himbert, M. E.},
  title = {Improved frequency measurement and isotope shift of the $4\mathrm{d}^95\mathrm{s}^2~^2\mathrm{D}_{5/2} \rightarrow 4\mathrm{d}^{10}6\mathrm{p}~^2\mathrm{P}_{3/2}$ transition in silver by laser heterodyne spectroscopy},
  journal = {Eur. Phys. J. D},
  volume = {14},
  pages = {39-42},
  year = {2001},
  doi = {10.1007/s100530170232},
}

@article{Badr2004,
  author = {Badr, T. and Plimmer, M. D. and Juncar, P. and Himbert, M. E. and Silver, J. D. and Rovera, G. D.},
  title = {Continuous-wave {D}oppler-free two-photon spectroscopy of the $4\mathrm{d}^{10}5\mathrm{s}~^2\mathrm{S}_{1/2} \rightarrow 4\mathrm{d}^{9}5\mathrm{s}^2~^2\mathrm{D}_{3/2}$ transition in atomic silver},
  journal = {Eur. Phys. J. D},
  volume = {31},
  pages = {3-10},
  year = {2004},
  doi = {10.1140/epjd/e2004-00118-y},
}

@article{Badr2006,
  title = {Observation by two-photon laser spectroscopy of the $4{d}^{10}5s\phantom{\rule{0.2em}{0ex}}^{2}\mathrm{S}_{1/2}\ensuremath{\rightarrow}4{d}^{9}5{s}^{2}\phantom{\rule{0.2em}{0ex}}^{2}\mathrm{D}_{5/2}$ clock transition in atomic silver},
  author = {Badr, T. and Plimmer, M. D. and Juncar, P. and Himbert, M. E. and Louyer, Y. and Knight, D. J. E.},
  journal = {Phys. Rev. A},
  volume = {74},
  pages = {062509},
  year = {2006},
  doi = {10.1103/PhysRevA.74.062509},
}

@article{Bengyel2026,
  title = {More detail of the laser system are to be given in a future manuscript that is in preparation which focuses on the magneto-optical trap apparatus.},
}

@article{Dahmen1967,
  title = {Measurements of the nuclear magnetic dipole moment of {Au$^{197}$} and Hyperfine structure measurements in the ground states of {Au$^{197}$}, {Ag$^{107}$} and {K$^{39}$}},
  author = {Dahmen, H. and Penselin, S.},
  journal = {Z. Physik},
  volume = {200},
  pages = {456-466},
  year = {1967},
  doi = {10.1007/BF01326186},
}

@article{Carlsson1990,
  title = {Accurate time-resolved laser spectroscopy on silver atoms},
  author = {Carlsson, J. and J\"{o}nsson, P. and Sturesson, L.},
  journal = {Z. Phys. D - Atoms, Molecules and Clusters},
  volume = {16},
  pages = {87-90},
  year = {1990},
  doi = {10.1007/BF01679568},
}

@article{Pickering2001,
  title = {New accurate data for the spectrum of neutral silver},
  author = {Pickering, J. C. and Zilio, V.},
  journal = {Eur. Phys. J. D},
  volume = {13},
  pages = {181-185},
  year = {2001},
  doi = {10.1007/s100530170264},
}

@article{Sherlock2009,
  title = {How weak is a weak probe in laser spectroscopy},
  author = {Sherlock, Ben E. and Hughes, Ifan G.},
  journal = {Am. J. Phys.},
  volume = {77},
  pages = {111-115},
  year = {2009},
  doi = {10.1119/1.3013197},
}

@article{Pitz2014,
  title = {Pressure broadening and shift of the rubidium $D_1$ transition and potassium $D_2$ transitions by various gases with comparison to other alkali rates},
  author = {Pitz, Greg A. and Sandoval, Andrew J. and Tafoya, Tiffany B. and Klennert, Wade L. and Hostutler, David A.},
  journal = {J. Quant. Spectrosc. Radiat. Transf.},
  volume = {140},
  pages = {18-29},
  year = {2014},
  doi = {10.1016/j.jqsrt.2014.01.024},
}

@article{Loreau2013,
  title = {Potential energy curves for the interaction of {Ag}(5s) and {Ag}(5p) with noble gas atoms},
  author = {Loreau, J. and Sadeghpour, H. R. and Dalgamo, A.},
  journal = {J. Chem. Phys.},
  volume = {138},
  pages = {084301},
  year = {2013},
  doi = {10.1063/1.4790586},
}

@article{Mohr2015,
  title = {Dimensionless units in the SI},
  author = {Mohr, Peter J. and Phillips, William D.},
  journal = {Metrologia},
  volume = {52},
  pages = {40},
  year = {2015},
  doi = {10.1088/0026-1394/52/1/40},
}

@article{Foley1946,
  title = {The pressure broadening of spectral lines},
  author = {Foley H. M.},
  journal = {Phys. Rev.},
  volume = {69},
  pages = {616},
  year = {1946},
  doi = {10.1103/PhysRev.69.616},
}

@article{Mizushima1951,
  title = {The theory of pressure broadening and its application in microwave spectra},
  author = {Mizushima, Masataka},
  journal = {Phys. Rev.},
  volume = {83},
  pages = {94},
  year = {1951},
  doi = {10.1103/PhysRev.83.94},
}

@article{NIST,
  title = {National institute of standards and technology: {Atomic} spectra database 78},
  journal = {Accessed on 10th August 2026.},
  doi = {10.18434/T4W30F},
}

@article{luo2021thermal,
  title={Thermal and temporal characteristics of Faraday anomalous dispersion optical filters based on a hollow cathode lamp},
  author={Luo, Bin and Ma, Rui and Ji, Qianqian and Yin, Longfei and Chen, Jingbiao and Guo, Hong},
  journal={Opt. Lett.},
  volume={46},
  number={21},
  pages={5372--5375},
  year={2021},
  publisher={Optical Society of America}
}

@article{degen2017quantum,
  title={Quantum sensing},
  author={Degen, Christian L and Reinhard, Friedemann and Cappellaro, Paola},
  journal={Rev. Mod. Phys.},
  volume={89},
  number={3},
  pages={035002},
  year={2017},
  publisher={APS}
}

@article{fabricant2023build,
  title={How to build a magnetometer with thermal atomic vapor: a tutorial},
  author={Fabricant, Anne and Novikova, Irina and Bison, Georg},
  journal={New J. Phys.},
  volume={25},
  number={2},
  pages={025001},
  year={2023},
  publisher={IOP Publishing}
}

@article{downes2023practical,
  title={A practical guide to terahertz imaging using thermal atomic vapour},
  author={Downes, Lucy A and Torralbo-Campo, Lara and Weatherill, Kevin J},
  journal={New J. Phys.},
  volume={25},
  number={3},
  pages={035002},
  year={2023},
  publisher={IOP Publishing}
}

@article{pizzey2022laser,
  title={Laser spectroscopy of hot atomic vapours: from’scope to theoretical fit},
  author={Pizzey, D and Briscoe, JD and Logue, FD and Ponciano-Ojeda, FS and Wrathmall, SA and Hughes, IG},
  journal={New J. Phys.},
  volume={24},
  number={12},
  pages={125001},
  year={2022},
  publisher={IOP Publishing}
}

@book{thorne1999spectrophysics,
  title={Spectrophysics: principles and applications},
  author={Thorne, Anne and Litz{\'e}n, Ulf and Johansson, Sveneric},
  year={1999},
  publisher={Springer Science \& Business Media}
}

@article{lewis1980collisional,
  title={Collisional relaxation of atomic excited states, line broadening and interatomic interactions},
  author={Lewis, EL},
  journal={Phys. Rep.},
  volume={58},
  number={1},
  pages={1--71},
  year={1980},
  publisher={Elsevier}
}

@article{weller2011absolute,
  title={Absolute absorption on the rubidium D1 line including resonant dipole--dipole interactions},
  author={Weller, Lee and Bettles, Robert J and Siddons, Paul and Adams, Charles S and Hughes, Ifan G},
  journal={J. Phys. B: At. Mol. Opt. Phys.},
  volume={44},
  number={19},
  pages={195006},
  year={2011}
}

@book{hughes2010measurements,
  title={Measurements and their uncertainties: a practical guide to modern error analysis},
  author={Hughes, Ifan and Hase, Thomas},
  year={2010},
  publisher={Oup Oxford}
}

@article{alqarni2025role,
  title={The role of buffer gas in shaping the D1 line spectrum of potassium vapour},
  author={Alqarni, Sharaa A and Pizzey, Danielle and Wrathmall, Steven A and Hughes, Ifan G},
  journal={J. Phys. B: At. mol. Opt. Phys.},
  volume={58},
  number={13},
  pages={135201},
  year={2025},
  publisher={IOP Publishing}
}

@article{blums2020laser,
  title={Laser stabilization to neutral Yb in a discharge with polarization-enhanced frequency modulation spectroscopy},
  author={Bl{\=u}ms, Valdis and Scarabel, Jordan and Shimizu, Kenji and Ghadimi, Moji and Connell, Steven C and H{\"a}ndel, Sylvi and Norton, Benjamin G and Bridge, Elizabeth M and Kielpinski, David and Lobino, Mirko and others},
  journal={Rev. Sci. Instrum.},
  volume={91},
  number={12},
  year={2020},
  publisher={AIP Publishing}
}

@article{sarmiento2018comparing,
  title={Comparing the emission spectra of U and Th hollow cathode lamps and a new U line list},
  author={Sarmiento, LF and Reiners, A and Huke, P and Bauer, FF and Guenter, EW and Seemann, U and Wolter, U},
  journal={A\&A},
  volume={618},
  pages={A118},
  year={2018},
  publisher={EDP Sciences}
}

@article{liu2018ultraviolet,
  title={Ultraviolet laser spectroscopy of aluminum atoms in hollow-cathode lamp},
  author={Liu, Hongli and Yuan, Wenhao and Cheng, Feihu and Wang, Zhiyuan and Xu, Zetian and Deng, Ke and Lu, Zehuang},
  journal={J. Phys. B: At. Mol. Opt. Phys.},
  volume={51},
  number={22},
  pages={225002},
  year={2018},
  publisher={IOP Publishing}
}

@book{sobelman2012atomic,
  title={Atomic spectra and radiative transitions},
  author={Sobelman, Igor I},
  year={2012},
  publisher={Springer Science \& Business Media}
}

@article{Bengtsson1990,
  title = {Hyperfine structure and radiative-lifetime determination for the 4${\mathit{d}}^{10}$6p $^{2}$P states of neutral silver using pulsed laser spectroscopy},
  author = {Bengtsson, J. and Larsson, J. and Svanberg, S.},
  journal = {Phys. Rev. A},
  volume = {42},
  issue = {9},
  pages = {5457--5463},
  numpages = {0},
  year = {1990},
  month = {Nov},
  publisher = {American Physical Society},
  doi = {10.1103/PhysRevA.42.5457},
  url = {https://link.aps.org/doi/10.1103/PhysRevA.42.5457}
}

\section*{Supplementary Material}
\subsection*{Transition strength coefficients in an applied magnetic field}

The presented work is done in the absence of a magnetic field. The addition of a magnetic field would split each state $F$ into $(2F+1)$ hyperfine Zeeman sublevels that may be labelled by the projection $m_F$ of $F$ along the field direction. The transition strength coefficients in the magnetic field $c_{m_F}$ then depend upon $F, m_F, F', m_{F'}$. The values of $c_{m_F}$ can be calculated using the relation
\begin{equation}
\nonumber
c_{m_F}=c_{F}\times
\begin{pmatrix}
F' & 1 & F \\
m_{F'} & -q & -m_F
\end{pmatrix}
\end{equation}
where the term in brackets is a Wigner 3-$j$ coefficient, and $q$ depends on the polarisation of the light and is equal to the the integer change in $m_F$ during the transition. We give the relevant values of $c_{m_F}$ for the $5{^{2}}\mathrm{S}_{1/2} \rightarrow 5{^{2}}\mathrm{P}_{3/2}$ of silver in the table below: 
\\
\begin{table}[h!]
\centering
\begin{tabular}{cc|cccccccc}
 & & \multicolumn{8}{c}{$(F', m_{F'})$} \\
 & & (1,-1) & (1,0) & (1,+1) & (2, -2) & (2, -1) & (2, 0) & (2, +1) & (2, +2) \\
 \hline
 \multirow{5}{*}{\rotatebox{90}{$(F, m_F)$}} & (0,0) & $\sqrt{2/9}$ & $\sqrt{2/9}$ & $\sqrt{2/9}$ & - & - & - & - & - \\
 & & & & \\
 & (1,-1) &  $\sqrt{1/18}$ & $\sqrt{1/18}$ & 0 & $\sqrt{1/3}$ & $\sqrt{1/6}$ & $\sqrt{1/18}$ & 0 & 0 \\
  & (1,0) & $\sqrt{1/18}$ & 0 & $\sqrt{1/18}$ & 0 & $\sqrt{1/6}$ & $\sqrt{2/9}$ & $\sqrt{1/6}$ & 0\\
   & (1,+1) & 0 & $\sqrt{1/18}$ & $\sqrt{1/18}$ & 0 & 0 & $\sqrt{1/18}$ & $\sqrt{1/6}$ & $\sqrt{1/3}$ \\
\end{tabular}
\end{table}

\end{document}